\documentclass[pra,nopacs,twocolumn,nofootinbib]{revtex4-1}
\usepackage{graphicx}
\setkeys{Gin}{draft=false} 
\usepackage{bm,amssymb,amsmath}
\usepackage{dcolumn}
\usepackage{natbib}
\usepackage
{hyperref}

\newcommand{\eEDM}{{\em e}EDM}

\newcommand{\Eref}[1]{Eq.~(\ref{#1})}

\begin{document}
\title{Manifestations of nuclear CP-violation in ThO molecule}

\author{L.V.\ Skripnikov$^{1,2}$}\email{leonidos239@gmail.com}
\author{A.N.\ Petrov$^{1,2}$}
\author{A.V.\ Titov$^{1,2}$}
\homepage{http://www.qchem.pnpi.spb.ru}

   \affiliation{$^{1}$B.P.Konstantinov Petersburg Nuclear Physics Institute, Gatchina, Leningrad district 188300, Russia}
\affiliation{$^{2}$Dept.\ of Physics, Saint Petersburg State University, Saint Petersburg, Petrodvoretz 198504, Russia}

\author{V. V. Flambaum$^{3}$}
\affiliation{$^3$School of Physics, The University of New South Wales, Sydney
NSW 2052, Australia}

\date{\today, 20t}

\begin{abstract}
Investigations of CP violation in hadron sector may be done using measurements in the ThO molecule. Recent measurements in this molecule improved the limit on electron EDM by an order of magnitude. Another time reversal (T) and parity (P) violating effect in $^{229}$ThO is induced by the nuclear magnetic quadrupole moment. We have performed nuclear and molecular calculations to express this effect in terms of the strength constants of T,P-odd nuclear forces,
neutron EDM, QCD vacuum angle $\theta$, quark EDM and chromo-EDM.
\end{abstract}

\maketitle

\section{Introduction}

The best limits on the electric dipole moment (EDM) of the proton and T,P-violating nuclear forces have been obtained using the measurements of Hg atom EDM \cite{HgPRL,Hg}. The Hg EDM measurements also give a limit on the neutron EDM, which is only twice weaker than that from the direct neutron EDM measurement.  There are also measurements in other diamagnetic atoms (Xe, Ra, Rn) \cite{Hg,Xe,Ra,Rn} and TlF molecule \cite{Cho:91, Petrov:02}.

The problem is that the nuclear EDM, $d_N$, in neutral atoms and molecules is screened by electrons (the Schiff theorem) and can not  be measured directly.
Therefore, atomic EDM in heavy  diamagnetic atoms is generated by the Schiff moment \cite{FKS84b,Sandars}. The Schiff moment is $\sim r_N^2 d_N$, where $r_N$ is a very small nuclear radius \cite{Flambaum:02} on the atomic scale.
As a result, the atomic EDM produced by the nuclear Schiff moment is significantly smaller than the nuclear EDM.

The magnetic interaction between the nuclear moments and electrons is not screened. The lowest T,P-odd magnetic moment is the nuclear magnetic quadrupole moment (MQM). To have MQM working we should consider paramegnetic atoms and molecules, where the electron angular momentum is not zero, and electrons produce a magnetic field interacting with MQM. It was shown in Ref.~\cite{FKS84b} that in paramagnetic atoms and molecules MQM induces
larger EDM than the Schiff moment (see also \cite{GFreview,Khr91}). Also, it was shown in Ref.~\cite{F94}  that in deformed nuclei MQM has a collective nature and is significantly enhanced (remind the reader that an ordinary electric quadrupole moment is also enhanced in deformed nuclei). Remarkably, in all molecules of current experimental interest a heavy atom has a deformed nucleus and this collective enhancement works (in isotopes with nuclear spin $I>1/2$, where MQM exists). 

Refs.~\cite{FKS84b,KL95,FDK14} suggested to use paramagnetic molecules to measure T,P-violating effects produced by MQM. Heavy diatomic molecules with  $^3\Delta_1$ electron term look especially promissing \cite{FDK14}. There are several reasons. Firstly, the effect of MQM very rapidly increases with the nuclear charge $Z$ \cite{FKS84b}. Second,  the $^3\Delta_1$ electron term has $\Omega$-doublet structure with a very small interval between the opposite parity levels. This allows one to polarize the molecule by a weak electric field and cancel some systematic errors since the effect on the doublet components has an opposite sign \cite{SF78,ComminsFest,Petrov:14}. Magnetic moment of the $^3\Delta_1$ electron term is very small, and this is another reason for reducing the systematic errors.  Finally, a new experimental technique was developed which allowed one to improve the limit on electron electric dipole moment using ThO molecule by more than an order of magnitude \cite{ThO}.

The aim of the present paper is to perform accurate calculations of MQM effect in ThO which should allow one to measure nuclear CP-violating interactions and nucleon EDM using ThO experiments \cite{ThO}. These measurements provide a method to search for physics beyond the Standard model and test unification theories.

The T,P-odd electromagnetic interaction of the nuclear magnetic quadrupole moment with electrons is described by the Hamiltonian \cite{Kozlov:87}:
 \begin{align}\label{hamq}
 H  &=
 -\frac{  M}{2I(2I-1)}  T_{ik}\frac{3}{2r^5}\epsilon_{jli}\alpha_jr_lr_k,
 \end{align}
where $\epsilon_{jli}$ is the unit antisymmetric tensor, $\bm{\alpha}$ is the vector of Dirac matrices, $\bm{r}$ is the displacement of the electron from the Th nucleus, $\bm I$ is the nuclear spin,  $M$ is the nuclear MQM,
\begin{align}\label{eqaux1}
M_{i,k}=\frac{3M}{2I(2I-1)}T_{i,k}\, \\
 T_{i,k}=I_i I_k + I_k I_i -\tfrac23 \delta_{i,k} I(I+1)\,.
 \end{align}
In the subspace of the $\pm \Omega$ states Hamiltonian (\ref{hamq}) is reduced to 
the following effective molecular Hamiltonian \cite{FKS84b}:
 \begin{align}\label{eq0}
 H_\mathrm{eff} &=
 -\frac{W_M  M}{2I(2I-1)} \bm S \hat{\bm T} \bm n
 \,.
 \end{align}
Here parameter $W_M$, \Eref{WM}, will be found from the molecular calculations, $\bm S$ is the effective electron  spin ~\cite{KL95}, $S{=}|\Omega|{=}1$, $\bm n$ is a unit vector directed along the molecular axis $\zeta$ from Th to O,
$\Omega= {\bm J^e \cdot \bm{n}}$ is the projection of the total electronic angular momentum ${\bm J^e} $ on the molecular axis.
Note, that contrary to the $|\Omega|=1/2$ case, Hamiltonians (\ref{hamq},\ref{eq0}) do not mix $\Omega=\pm1$ components.
Neglecting the interaction between different rotational levels one can
obtain that MQM energy shift is
\begin{eqnarray}
\label{shift}
\delta(J,F) = (-1)^{\Omega+I+F+1}C(J,F) W_M M\ , \\
C(J,F)= \frac{(2J+1)}{2}
\frac{
    \left(
    \begin{array}{ccc}
    J &  2 &  J \\
   -\Omega & 0 & \Omega
    \end{array}
    \right)
    }
    {
    \left(
    \begin{array}{ccc}
    I &  2 &  I \\
   -I & 0 & I
    \end{array}
    \right)
    }
    \left\{
    \begin{array}{ccc}
    J &  I &  F \\
    I &  J & 2
    \end{array}
    \right\},
\end{eqnarray}
where F is the total angular moment.
For $^{229}$ThO ($I=5/2$) and ground rotational level {$J{=}1$} \Eref{shift} gives
MQM energy shifts, 
$\left| \delta(J,F) \right|$,
equal to $0.14 W_M M, \ 0.16 W_M M, \ 0.05 W_M M$ for 
$F= 3/2,5/2,7/2$, correspondingly.


\section{Nuclear Magnetic Quadrupole moment}

    Using the results of \cite{FKS84b,KhMQM},  MQM of a valence nucleon in a spherical nucleus  may be presented as  
\begin{align}\label{M}
M=[d-2\cdot 10^{-21} \eta(\mu-q)  (e \cdot cm)] \lambda_p (2I-1)t_I,
 \end{align}
where $t_I=1$ for $I=l+1/2$ and $t_I=-I/(I+1)$ for $I=l-1/2$, $I$ and $l$ are the total and orbital angular momenta of a valence
nucleon, $\eta$ is the dimentionless strength constant of the T,P-odd nuclear potential $\eta G/(2^{3/2} m_p) (\sigma \cdot \nabla \rho)$ acting on the valence nucleon,  and $\rho$ is the total nucleon number density, the nucleon magnetic moments are $\mu_p=2.79$ for valence proton  and $\mu_n=-1.91$ for valence neutron, $q_p=1$ and $q_n=0$, $d$ is the valence nucleon EDM, $\lambda_p=\hbar /m_pc$. 
The MQM of a deformed nucleus in the ``frozen'' frame (rotating together with a nucleus) is given by the formula \cite{F94}:
 \begin{align}\label{Mzz}
M^{\rm nucl}_{zz}=\sum M^{\rm single}_{zz} (I,I_z,l) n(I,I_z,l),
 \end{align}
where $M^{\rm single}_{zz}(I,I_z,l)$ is given by Eqs.~(\ref{M}) and (\ref{eqaux1}), $T_{zz}=2 I_z^2 -\tfrac23 I(I+1)$, $n(I,I_z,l)$ are the orbital occupation numbers.  
For $^{229}$Th nucleus the occupation numbers have been found using the diagrams presented in Ref.~\cite{Bohr}: 13 neutrons on the orbitals  $g_{9/2}$, $I_z= 5/2,\pm 3/2, \pm 1/2$;
$j_{15/2}$, $I_z= \pm5/2,\pm 3/2, \pm 1/2$; 
$i_{11/2}$, $I_z=  \pm 1/2$; and 6 protons on the orbitals $h_{9/2}$, $I_z= \pm 3/2, \pm 1/2$; $i_{13/2}$, $I_z=  \pm 1/2$.

 The MQM in the laboratory frame $M\equiv M_{\rm lab}$ can be expressed via MQM in the rotating frame (\ref{Mzz}):
\begin{align}\label{Mlab}
  M^{\rm lab}=\frac{I(2I-1)}{(I+1)(2I+3)} M^{\rm nucl}_{zz}=\\
[7\cdot 10^{-20} \eta_n \mu_n  (e \cdot {\rm cm}) -19 d_n] \lambda_p \,,
 \end{align}
where 
  $I=5/2$ 
is the $^{229}$Th nuclear spin.
The proton contribution is small due to an accidental cancellation of the contributions of different orbitals.

The T,P-odd nuclear forces are dominated by the $\pi_0$ meson exchange. Therefore, we may express the strength constants via strong $\pi NN$ coupling constant $g=13.6$ and T,P-odd $\pi N N$ coupling constants corresponding to the isospin channels  $T=0,1,2$:  $\eta_n=  5\cdot 10^6 g (  {\bar g_1}{+}0.4{\bar g_2}{-}0.2{\bar g_0}) $. The numerical coefficient here was obtained as a product of two factors:  $[Gm_{\pi}^2/2^{1/2} ]^{-1}=6.7 \cdot 10^{6}$ from the $\pi-$meson exhange in the zero-range  limit and  the factor 0.7 corresponding to the zero-range reduction of the finite range interaction due to the $\pi_0-$exchange  \cite{FKS84b,DKT}.
For the charge meson exchange the reduction factor is about 0.16 (since the exchange interaction contains a small overlap of the proton and neutron wave functions in the $\pi^-pn$ vertex), and we neglect this contribution.  We have also included two additional correction factors for the value of $M$. More accurate numerical calculations in Saxon-Woods potentail \cite{FKS84b,DKT} give larger values  of MQM  (the factor $\sim$ 1.2) than the simple analytical solution  in Eq.~(\ref{M}), on the other hand, the many-body corrections reduce the effective strength constants of
  T,P-odd potential $\eta$ $\sim$1.5 times \cite{F94,FVeta}.
As a result, we obtain  
\begin{align}\label{Mg}
\nonumber 
M(g)= - [g (  {\bar g_1}+ 0.4 {\bar g_2}-0.2 {\bar g_0}) \\
+d_n/(1.4 \cdot 10^{-14}  e \cdot cm)] \cdot  6\cdot 10^{-27} e \cdot cm^2 .
 \end{align}
Possible CP-violation in the strong interaction sector is described by the  CP violation parameter ${\tilde \theta}$. According to Ref.~\cite{theta}  $g {\bar g_0}=-0.37 {\tilde \theta} $. This gives the following value of MQM for $^{229}$Th:
\begin{align}\label{Mtheta}
M(\theta) = -4\cdot 10^{-28} {\tilde \theta} \cdot e \cdot cm^2 .
 \end{align} 
Finally, we can express MQM in terms of the quark chromo-EDM ${\tilde d_u}$ and  ${\tilde d_d}$ using the relations 
 $g {\bar g_1}=4.{\cdot}10^{15}( {\tilde d_u} - {\tilde d_d})/cm $, $g {\bar g_0}=0.8 \cdot 10^{15}( {\tilde d_u} + {\tilde d_n})/cm $ 
 \cite{PospelovRitzreview}:
\begin{align}\label{Md}
 M( {\tilde d}) = -2 \cdot 10^{-11} ( {\tilde d_u} - {\tilde d_d}) \cdot e \cdot {\rm cm} .
 \end{align}
The contributions of $d_n$ to MQM  in Eqs.~(\ref{Mg} -\ref{Md}) are from one to two orders of magnitude smaller than the contributions of the nucleon CP-odd interactions.

\

\section{Electronic structure calculation}

 To obtain $W_M$ in the ThO molecule theoretically, one can evaluate the following matrix element \cite{FDK14}:
\begin{align}\label{WM}
W_M= \frac{3}{2}\frac{1}{\Omega} 
   \langle
   \Psi_{^{3}\Delta_1}\vert\sum_i\left(\frac{\bm{\alpha}_i\times
\bm{r}_i}{r_i^5}\right)
 _\zeta r_\zeta \vert\Psi_{^{3}\Delta_1}
   \rangle,
 \end{align}
where $\Psi$ is the electronic wave function of the considered ThO state.

The matrix element (\ref{WM}) is a mean value of the operator heavily concentrated in the atomic core of Th and sensitive to variation of core-region spin densities of the valence electrons. 
This is example of the so-called ``atom in a compound'' or AiC properties \cite{Titov:14}.
Efficient and very accurate computations of such properties can be performed by a two-step approach \cite{Titov:06amin, Petrov:02} utilizing the generalized relativistic effective core potential (GRECP) method \cite{Titov:99, Mosyagin:10a}. In the first (molecular) step the GRECP is used to exclude the inner-core electrons from a correlation calculation and obtain an accurate description of the valence part of the wave function by an economical way. Thus, the computational cost of the relativistic molecular calculation is dramatically reduced. 
It should be noted that the GRECP operator allows one to take account of the Breit interaction very effectively \cite{Petrov:04b, Mosyagin:06amin}.
Second, a nonvariational restoration procedure is employed \cite{Titov:06amin} to recover the valence wave function in the inner core region of a heavy atom.
The procedure is based on a proportionality of valence and virtual spinors in the inner-core regions of heavy atoms. To perform the restoration one generates {\it equivalent} basis sets of one-center four-component spinors 
$$
  \left( \begin{array}{c} f_{nlj}(r)\theta_{ljm} \\
     g_{nlj}(r)\theta_{2j{-}l,jm} \\ \end{array} \right)
$$
 and smoothed two-component pseudospinors
$$
    \tilde f_{nlj}(r)\theta_{ljm}
$$
in all-electron finite-difference Dirac-Fock-Breit and GRECP~/~self-consistent field calculations (employing the $jj-$coupling scheme) of {the same} configurations of a considered atom and its ions \cite{HFDB, Bratzev:77, HFJ, Tupitsyn:95}. These sets, describing mainly the given atomic core region, are generated independently of the basis set exploited in the molecular GRECP calculations.
A first order reduced density matrix obtained at the first step is reexpanded into the basis of smoothed two-component pseudospinors.
Replacing these pseudospinors by equivalent four-component spinors one obtains the true four-component density matrix. Taking trace of the product of the density matrix with the matrix form of an operator describing a given property
one obtains the expectation value of the property. Note that the numerical form of four-component spinors is used which allows one to get a correct form of the wavefunction in the core region of a given heavy atom.

The single-reference two-component relativistic coupled-clusters method with single, double and perturbative treatment of triple cluster amplitudes, CCSD(T), was used to take account of both the relativistic and correlation effects for valence electrons.
The $1s-4f$ inner-core electrons of Th were excluded from the molecular correlation calculations using the ``valence'' semi-local version of the GRECP operator \cite{Mosyagin:10a}. Thus, 38 electrons ($5s^2 5p^6 5d^{10} 6s^2 6p^6 6d^2 7s^2 $ (Th) and $1s^2 2s^2 2p^4$(O))
were treated explicitly in our correlation calculations.

A basis set for Th from Ref.~\cite{Skripnikov:13c} was used with extended number of $d$ functions.  It can be designated as (30,20,10,11,4,1)/[30,8,10,4,4,1].
For oxygen the aug-ccpVQZ basis set \cite{Kendall:92} with removed two g-type basis functions was employed, i.e., we used the (13,7,4,3)/[6,5,4,3] basis set.

Within the (G)RECP approach it is possible to exclude the spin-orbit effects for valence electrons only and, thus, perform the scalar-relativistic calculations \cite{Mosyagin:10a}. 
This leads to considerable computational savings and allows one to use larger basis sets  exploiting the same computational resources. 
We used this way to calculate the correction for $W_M$ on the basis set enlargement.
For this we have performed: {\it (i)} scalar-relativistic CCSD(T) calculation using the same basis set as used for the two-component calculation; {\it (ii)} scalar-relativistic CCSD(T) calculation using the extended basis set on Th [22,17,15,14,10,10,5]\footnote{The number of $s$ basis functions was reduced  because of a linear dependence problem, however, it did not influence the accuracy of the evaluated property.} and extended basis set on O -- aug-ccpCVQZ basis set \cite{Kendall:92} with removed $g$-type basis functions, (16,10,6,4)/[9,8,6,4].
The corrections were estimated as differences between the values of the corresponding parameters. 
Note that {\it no cuts} of the active space of orbitals by energy were done in the correlation studies, i.e., in the 38-electron CCSD(T) calculation with the largest basis set in which all 1204 spin-orbitals were involved in the calculation explicitly.

The experimental equilibrium internuclear distance \cite{Huber:79,Edvinsson:84}  3.511 a.u.\ for $H^3\Delta_1$ state was used in these calculations. It was shown in \cite{Skripnikov:13c} that the calculated equilibrium internuclear distance as well as harmonic frequencies are very close to the experimental data \cite{Huber:79,Edvinsson:84}.

The coupled-clusters calculations were performed using the {\sc dirac12} \cite{DIRAC12}, {\sc mrcc} \cite{MRCC2013} and {\sc cfour} codes \cite{CFOUR}. The nonvariational restoration code developed in \cite{Skripnikov:13b, Skripnikov:13c, Skripnikov:11a} and interfaced to these codes was used to restore the four-component electronic structure near the Th nucleus. The expectation value of the operator corresponding to $W_M$ (\ref{WM}) was calculated using the code developed in the present paper.

\begin{table}[!h]
\caption{
The calculated values of $W_M$ parameter of the $H ^3\Delta_1$ state of ThO 
using the coupled-clusters methods.
The GRECP calculations were performed with (1c) and without (2c) accounting for the spin-orbit effects.
}
\label{TResultsWm}
\begin{tabular}{ l  c  c  c  c  c}
\hline\hline
Method & $W_M, \frac{10^{33}Hz}{e~{\rm cm}^2}$  \\
\hline       
  1c-CCSD      & {} 1.81  \\  
  1c-CCSD(T)   & {} 1.76  \\       
\hline
  2c-CCSD      & {} 1.74  \\  
  2c-CCSD(T)   & {} 1.68  \\   
\hline  
  2c-CCSD(T)   & {} 1.66  \\
+ basis corr.  & {}       \\
(Final)  & {}       \\ 
\hline\hline
\end{tabular}
\end{table} 

The result of earlier performed ``semiempirical'' estimate \cite{FDK14}, 1.9 $\frac{10^{33}Hz}{e~{\rm cm}^2}$, is in a good agreement with the current {\it ab~initio} calculations
\footnote{Such a coincidence is not accident. Since there were no experimental data obtained for the hyperfine structure constants of the $H ^3\Delta_1$ state of ThO up to now, the given ``semiempirical'' value is obtained on the basis of our previous {\it ab~initio} calculation of an effective electric field \cite{Skripnikov:13c},  and the current and previous our {\it ab~initio} calculations are very close methodologically.}.

It follows from Table~\ref{TResultsWm} that the spin-orbit contribution to $W_M$ is $-0.08$ $\frac{10^{33}Hz}{e~{\rm cm}^2}$.

According to the density matrix analysis the main contribution to $W_M$ comes from mixing of $s$ and $p$ orbitals of Th while contribution from mixing $p$ and $d$ orbitals of Th is negligible.

Exclusion of 20 outer core electrons ($1s$(O), $5s^2 5p^6 5d^{10}$ (Th)) from the correlation treatment reduces $W_M$ value by 0.1 $\frac{10^{33}Hz}{e~{\rm cm}^2}$ that is not a small value, therefore, correlation of these electrons should taken into account in accurate calculation.

In the scalar-relativistic CCSD(T) calculations we have found that $W_M$ only very slightly depends on the internuclear distance, it decreases monotonically by 0.004~$\frac{10^{33}Hz}{e~{\rm cm}^2}$ from R(Th--O)=3.4~a.u.\ to R(Th--O)=3.56~a.u. 

In the two-component CCSD(T) calculations we used the orbitals from the two-component Hartree-Fock calculation of the $^1\Sigma^+$ state. Thus, the reference determinant for the coupled-clusters calculation of the $^3\Delta_1$ state was constructed from this set of spinors. 
To estimate the uncertainty of $W_M$ due to a particular choice of the reference determinant we have performed the following three scalar-relativistic calculations: {\it (i)}  using the orbitals (both in the reference determinant and in excited configurations) obtained from the restricted Hartree-Fock calculation of the $^1\Sigma^+$ state (like that in the two-component calculation), this corresponds to the so-called QRHF-reference \cite{Bartlett:95}, {\it (ii)} using restricted open-shell Hartree-Fock (ROHF) orbitals for the $^3\Delta$ state, and {\it (iii)} using unrestricted open-shell Hartree-Fock (UHF) orbitals for the $^3\Delta$ state.
The results coincide within 0.02 $\frac{10^{33}Hz}{e~cm^2}$, i.e., the uncertainty due to a particular choice of the reference configuration
can be estimated as  1\%. 
Such a weak dependence on the orbitals choice is due to a well-known advantage
of the coupled-clusters method when single-particle cluster amplitudes are included to the exponential ansatz, because of its ability to take account of the effects of orbital relaxation efficiently (see, e.g., \cite{Bartlett:95}). 
  Analyzing these results as well as the results from Table \ref{TResultsWm} and our earlier studies within the two-step procedure (e.g., see \cite{Lee:13a}) 
we expect that the theoretical uncertainty for our final value of the $W_M$ parameter is smaller than 7\%.

Finally, one can express the MQM energy shift, $C(J,F)\ W_M M$, in terms of the fundamental CP-violating physical quantities $\tilde{\theta}$ and $\tilde{d}_{u,d}$ using Eqs.~(\ref{Mtheta},\ref{Md}).
For the largest coefficient, $C(J{=}1,F{=}5/2)=0.16$, we have
\begin{align}
\label{EshiftTheta}
0.16 W_M  M  &= 
 -11 \cdot 10^{10} {\tilde \theta} \cdot \mu{\rm Hz} 
\end{align}  
\begin{align} 
\label{EshiftD} 
0.16 W_M  M  &=  
 -5\cdot \frac{10^{27}({\tilde d_u}-{\tilde d_d})}{\mathrm{cm}} \cdot \mu{\rm Hz}
\end{align} 
The current limits on $|{\tilde \theta}|$ and $|{\tilde d_u}{-}{\tilde d_d}|$ ($|{\tilde \theta}| < 2.4 \cdot 10^{-10}$, $|{\tilde d_u}{-}{\tilde d_d}|<6 \cdot 10^{-27} $~cm, see Ref.~\cite{Hg}) correspond to the shifts $|0.16\ W_M M | < 26~\mu$Hz and $30~\mu$Hz, respectively. 
The current accuracy in measurements of the energy shift produced by the \eEDM\ in $^{232}$ThO is $700~\mu$Hz \cite{ThO}. However, it is anticipated that it can be considerably improved by as much as $\sim\!2$ orders of magnitude \cite{ACME_Improvements}. Therefore, if one performs similar experiment on $^{229}$ThO the values of the frequency shifts produced by the nuclear MQMs are sufficiently large to compete in the improvement of limits on  the ${\tilde \theta}$-term and on the difference of the quark
chromo-EDMs $({\tilde d_u}-{\tilde d_d})$.

\section*{Acknowledgement}
The molecular calculations were partly performed on the Supercomputer ``Lomonosov''.
L.S., A.T.\ and A.P.\ acknowledge support from 
Saint Petersburg State University, research grant 0.38.652.2013 and 
RFBR Grant No.~13-02-01406. L.S.\ is also grateful to the President of Russian Federation grant no 5877.2014.2. V.F.\ acknowledges support from Australian Research Council and Humboldt Research Award. He is grateful to MBN Research Center for hospitality.


\end{document}